\documentclass[aps,pra,twocolumn,groupedaddress,nofootinbib,10pt]{revtex4-1}
\usepackage[english]{babel}
\usepackage{amsmath,amssymb,graphicx,enumerate,bbm,xcolor,latexsym}
\usepackage{mathtools}
\usepackage{comment}
\usepackage{paralist, tabularx}
\usepackage{nicefrac}
\usepackage{hyperref}
\newcommand{\rob}[1]{{\color{black} #1}}

\newcommand{\ket}[1]{\left| #1 \right>}

\newcommand{\beq}{\begin{equation}}
\newcommand{\eeq}{\end{equation}}

\usepackage{braket}

\newcommand{\LOSR}[0]{\ifmmode\textup{\upshape LOSR}\else{\textup{\upshape LOSR}}\fi}
\newcommand{\LO}[0]{\ifmmode\textup{\upshape LO}\else{\textup{\upshape LO}}\fi}
\newcommand{\LOCC}[0]{\ifmmode\textup{\upshape LOCC}\else{\textup{\upshape LOCC}}\fi}

\usepackage[only,Arrownot]{stmaryrd}

\begin{document}


\title{Experimentally adjudicating between different causal accounts of Bell inequality violations via statistical model selection}

\author{Patrick J. Daley, Kevin J. Resch}
\affiliation{Institute for Quantum Computing and Department of Physics \& Astronomy, University of Waterloo, Waterloo, Ontario N2L 3G1, Canada}
\author{Robert W. Spekkens}
\affiliation{Perimeter Institute for Theoretical Physics, 31 Caroline Street North, Waterloo, Ontario Canada N2L 2Y5}

\date{April 12, 2021}

\begin{abstract}
Bell inequalities follow from a set of seemingly natural assumptions about how to provide a causal model of a Bell experiment.  In the face of their violation, two types of causal models that modify some of these assumptions have been proposed:
(i) those that are parametrically conservative and structurally radical, such as models where the parameters are conditional probability distributions (termed `classical causal models') but where one posits inter-lab causal influences or superdeterminism, and (ii) those that are parametrically radical and structurally conservative, such as models where the labs are taken to be connected only by a common cause but where conditional probabilities are replaced by conditional density operators (these are termed `quantum causal models'). We here seek to adjudicate between these alternatives based on  their predictive power.
 The data from a Bell experiment  is divided into a training set and a test set, and for each causal model, the parameters that yield the best fit for the training set are estimated and then used to make predictions about the test set.   Our main result is that
   the structurally radical classical causal models 
    are disfavoured relative to the structurally conservative quantum causal model.  
  Their lower predictive power seems to be due to the fact that, unlike the quantum causal model, they are prone to a certain type of overfitting wherein statistical fluctuations away from the no-signalling condition are mistaken for real features.
     Our technique shows that it is possible to witness quantumness even in a Bell experiment that does not close the locality loophole. It also overturns the notion that it is impossible to experimentally test the plausibility of superdeterminist models of Bell inequality violations.
\end{abstract}

\maketitle



\textit{Introduction.}  There is now widespread agreement that the experimental evidence in favour of nature violating Bell inequalities~\cite{Bell64,CHSH,brunner2014bell}
 is persuasive~\cite{Belltest1,Belltest2,Belltest3}.  By contrast, 
 there is no agreement on how to provide a {\em causal account}
  of such violations~\cite{wood2015lesson}.
 One of the most popular views is that Bell inequality violations imply the existence of superluminal causal influences (typically understood as action at a distance)~\cite{maudlin2011quantum,norsen2006bell}.  Another view is that they imply the need for ``superdeterminism'', wherein  
the hidden variable that is a common cause of the two outcomes is also a cause of one or both of the setting variables
 ~\cite{sep-bell-theorem,t2016cellular,hossenfelder2020rethinking}.  A third option is that the correct causal account is one wherein there is just a common cause of the two outcomes, but wherein correlations are computed using the formalism of {\em quantum causal models}~\cite{LeiferSpekkens,allen2017quantum,CostaShrapnel,BLO}.
It is tempting to think that the only way to adjudicate between such accounts, i.e., the only way to assess the quality of the accounts of the correlations that they offer, is to consider their merits relative to some philosophical or aesthetic criteria. Recall, however, that this was the standard attitude towards the question of the merit
 of local hidden variable models of quantum theory until Bell 
showed how the question can be assessed by empirical data~\cite{Bell64}.  Motivated by Bell's example, we undertake to show that the selection problem articulated above can also be assessed empirically.

To date, proposed causal accounts of Bell inequality violations have typically been held to the following standard: {\em that the observations be reasonably likely given the model}. 
   In other words, a given causal model  has generally been considered unobjectionable on empirical grounds as long as it does not {\em underfit} the data.   From the perspective of statistical model selection, however, a more methodologically sound figure of merit when comparing models is their {\em predictive power}.  The latter can be compromised not only when the model underfits the data, but for other reasons as well, such as the model {\em overfitting} the data.
Overfitting occurs when the model-fitting procedure mistakes statistical fluctuations in the data for real features, a mistake which implies reduced predictive accuracy on novel data with different fluctuations.
In this article, we propose to 
hold causal accounts to a higher standard
by developing a technique for assessing the relative merit of different 
causal models using standard model selection tools. We then apply the technique to data obtained from a Bell-type experiment.\color{black}

In our comparison of alternative causal accounts of Bell inequality violations,
   we will focus on the distinction between a set of accounts    that are {\em structurally radical and parametrically conservative}, on the one hand, and an account that is   {\em structurally conservative and parametrically radical} on the other.  By ``structural radicalness,'' we mean that the particular {\em causal  structure} that appears in a causal account 
  is not the one that one would expect a priori to hold for a Bell scenario (more on this below). 
  By ``parametric radicalness,'' we mean that the mathematical formalism by which one extracts statistical predictions from a given structure is not the conventional one used in classical causal modelling,
  but rather the modification thereof proposed in the recent literature on {\em quantum causal models}~\cite{allen2017quantum,CostaShrapnel,BLO}.  The result of our analysis is that the latter sort of account is favoured by the experimental data.

In order to be able to compare the predictive power of different causal accounts
 of Bell inequality violations, we must cast them
  into a common framework.  For this purpose, we use a framework that subsumes both that
 of classical causal modelling~\cite{pearl2000causality,Spirtes00} and quantum generalizations thereof~\cite{allen2017quantum,CostaShrapnel,BLO}, thereby permitting   the relative predictive power of these views to be compared one to another using standard model selection techniques.
 Our proposal is an example of {\em causal discovery} using purely observational data.  Its relation to past work on causal discovery, both classical~\cite{pearl2000causality,Spirtes00} and quantum~\cite{giarmatzi2019quantum,bai2020efficient} is discussed in Appendix~\ref{priorwork}.

\color{black}

\begin{figure*}[t]
\begin{center}
\hspace{-2em}
\includegraphics[scale=0.25]{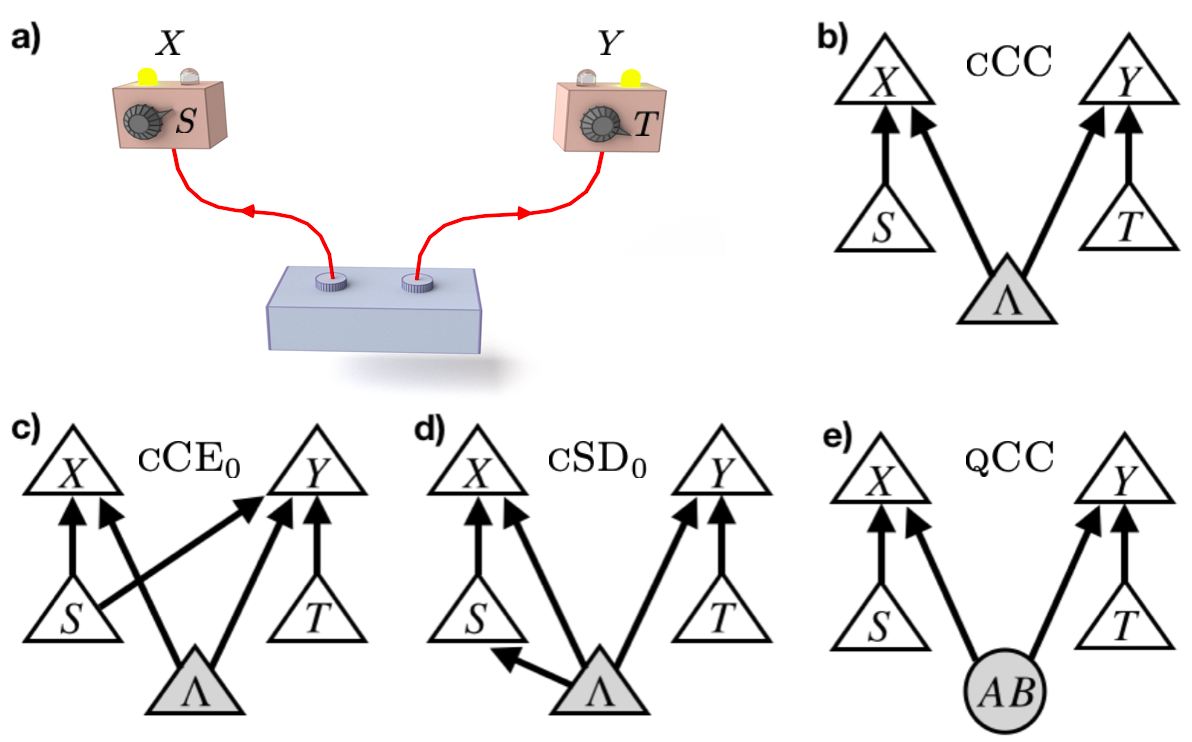}
\caption{(a) The bipartite Bell experiment and (b)-(e) the four causal models thereof that we consider here.
Triangular  nodes represent classical variables, while circular ones represent quantum systems; shaded nodes represent variables/systems that are latent.  
} 
\label{DAGs}
\end{center}
\vspace{-2em}
\end{figure*}

\textit{The framework of classical and quantum causal models.}
For classical causal models, we use Pearl's framework~\cite{pearl2000causality}.
The structural part of a classical causal model is a specification of 
the causal relations that hold among a set of systems (i.e., the {\em causal structure}) and is represented by a directed acyclic graph (DAG).  Examples are given in Figs.~\ref{DAGs}(b-d). Each node in the DAG represents a system, which in the classical case is associated to a random variable.
The directed edges into a node 
 $X$ from the causal parents of $X$, denoted ${\rm Pa}(X)$, represent 
 the potential for a direct causal influence in the interventional sense (namely, that manipulating a variable in ${\rm Pa}(X)$ while keeping all other variables fixed allows one to alter the statistical distribution of $X$).
The parametric part of a classical causal model stipulates, for every node $X$, 
\rob{the possibilities for } how the statistical distribution over $X$ \rob{can depend}
 on a given intervention on ${\rm Pa}(X)$, \rob{that is, the possibilities for the} 
conditional probability distribution  $P_{X|{\rm Pa}(X)}$, termed the ``do-conditional''.
\rob{A causal model may stipulate a restriction on the possibilities for $P_{X|{\rm Pa}(X)}$ for each node $X$, although in this article we are primarily interested in the case where there is no restriction, in which case we call the model {\em parameter-unrestricted}.}
Let $V$ denote the full set of variables in the DAG.  The distributions over $V$
  that are said to be {\em compatible} with the \rob{causal model}
 are those that can be expressed as the product $P_{V}=\prod_{X\in V} P_{X|{\rm Pa}(X)}$ 
 where $\{P_{X|{\rm Pa}(X)}\}_{X\in V}$ \rob{are conditional distributions in the set allowed by the model, which is {\em any} conditional distribution in the case of a parameter-unrestricted model.}
If only a subset, $O$, of the variables in $V$ are observed, such that the complementary set of variables, $V/O$, are unobserved (these are termed `latent variables'), then
 the compatible distributions on $O$ are computed by marginalization over the latent variables,
 $P_{O} = \sum_{V / O} P_{V}$. (Here, an expression such as $\sum_Y P_{XY}$ represents the distribution $P_X$ whose component at $X=x$ is $\sum_y P_{XY}(xy)$.)

Various proposals exist for how to define a quantum generalization of the notion of a causal model~\cite{allen2017quantum,CostaShrapnel,BLO}.  Although there are distinctions between these, they will not be relevant for the purposes of this article.
We follow Refs.~\cite{wood2015lesson, leifer2013towards} in taking the transition from classical to quantum causal models to be a transition in the nature of the parameters that supplement the causal structure, while the causal structure is taken to be
 represented in the same way as in a classical causal model, namely, by a DAG.
What one can \emph{infer} about a system $A$ given an intervention on its parents ${\rm Pa}(A)$ is no longer presumed to be represented by a conditional probability distribution,
 but is instead represented by a more exotic mathematical object, termed a {\em conditional density operator}, denoted $\rho_{A|{\rm Pa}(A)}$~\cite{leifer2013towards}. \rob{It is a positive operator on the tensor product of the Hilbert spaces of $A$ and ${\rm Pa}(A)$ that satisfies ${\rm Tr}_A (\rho_{A|{\rm Pa}(A)}) =\mathbb{I}_{{\rm Pa}(A)}$.  An equivalent way of representing what one can infer about $A$ given an intervention on ${\rm Pa}(A)$, which makes the analogy to conditional probabilities less obvious but connects better to the conventional formalism, is as a completely positive trace-preserving map from ${\rm Pa}(A)$ to $A$.  (The equivalence of these two representations is established using the Choi-Jamio{\l}kowski isomorphism.)}
\rob{The parameter-unrestricted versions of quantum causal models, which are the only ones we consider here, impose no restriction on the possibilities for the conditional density operator.
We will also focus here on quantum causal models wherein }
 the only systems that can be intrinsically quantum are the latent systems, while all of the observed systems are classical, since this simplifies the analysis and is sufficient to describe the Bell scenario (see Fig.~\ref{DAGs}(e)).

\textit{The slate of causal models--}
We will consider a bipartite Bell experiment, depicted in Fig.~\ref{DAGs}(a).
We refer to the two labs as `Alice's' and `Bob's'.
 The variable corresponding to the measurement setting (resp. outcome) in Alice's lab is denoted by $S$ (resp. $X$) and the variable describing the measurement setting (resp. outcome) in Bob's lab is denoted $T$ (resp. $Y$).
In the case we consider here, the outcome variables will be binary but the setting variables can take a larger number of values.
The conditional probability of outcomes given settings is denoted by $P_{XY|ST}$. 
The experimental data constitutes a finite sample from the distribution over $X$ and $Y$ for each set of values of $S$ and $T$, that is, a finite sample from $P^{(s,t)}_{XY} :=  \sum_{S,T} P_{XY|S,T} \delta_{S,s} \delta_{T,t}$ for each $(s,t)$.

%
We begin by describing the causal model that is excluded by 
Bell inequality violations.  As argued in Ref.~\cite{wood2015lesson}, this is the classical causal model with the DAG of Fig.~\ref{DAGs}(b), which 
describes a causal structure that fits Bell's intuitive notion of local causality~\cite{Bell64}, namely, that there is simply a common cause of the two outcomes. 
The correlations that are compatible with this causal model
 are those of the form~\cite{pearl2000causality}
\begin{align}
P^{{\rm \scalebox{.6}{\hspace{-.001cm}\textsc{cCC}}}}_{XY|ST} &:= \sum_{\Lambda} P_{X|S\Lambda} P_{Y|T\Lambda} P_{\Lambda},\label{clcmodel}
\end{align}
for some conditional probability distributions $P_{X|S\Lambda},P_{Y|T\Lambda},P_{\Lambda}$.
We refer to this model as  ``parametrically {\em classical} and structurally {\em common-cause}'', abbreviated \textsc{cCC}.
   It is the recasting, within the framework of causal models, of what is typically called a {\em local hidden variable model.} Distributions that are generated according to Eq.~(\ref{clcmodel}) satisfy Bell inequalities,
and therefore the \textsc{cCC} model is rejected if one observes a significant violation of such an inequality.  We now turn to a description of a variety of causal models that {\em can} violate Bell inequalities.

For anyone who presumes that the parameters in a causal model must be conditional probability distributions, it becomes necessary, in order to account for a Bell inequality violation, to presume a causal structure distinct from that of the DAG of Fig.~\ref{DAGs}(b).  We here consider the two most prominent classes of such parametrically conservative and structurally radical proposals.

 

The first class consists of those models that posit
  that there is a causal influence {\em from} the setting or outcome variable in one lab {\em to} the setting or outcome variable in the other lab, so that there is a cause-effect relation between the labs, rather than simply a common-cause relation.
We therefore refer to any such causal account  as parametrically {\em classical} and structurally {\em cause-effect}, abbreviated as \textsc{cCE}.
When the measurements in the labs are space-like separated---as they are in any experiment that seals the locality loophole~\cite{brunner2014bell}---these inter-lab causal influences must be superluminal\footnote{In fact, as shown in Ref.~\cite{bancal2014quantum}, such influences must have infinite speed.}.  
 In this article, we will consider one particular representative from the class of \textsc{cCE} models,
 corresponding to assuming the DAG depicted in Fig.~\ref{DAGs}(c).  We denote it by a subscript `0', i.e., \textsc{cCE}$_0$, simply as a reminder that there are other
 models in the \textsc{cCE} class.
We will refer to  the cardinality of the set of values that a variable $\Lambda$ can take as simply the {\em cardinality of $\Lambda$}.  We consider each possibility for the cardinality of $\Lambda$---up to the cardinality that saturates the set of achievable distributions~\cite{rosset2018universal}---as a distinct model.  (A more detailed discussion of the cardinality of $\Lambda$ is provided in Appendix~\ref{DetailsOnSlate}.)
 \rob{We assume that the model is parameter-unrestricted, so that 
we allow {\em any} classical parameter values.}
 
The compatible correlations in this case (for a fixed cardinality of $\Lambda$) are those that can be written as
\begin{align}
P^{_{\rm \scalebox{.6}{\hspace{-.001cm}cCE}_0}}_{XY|ST}&:= \sum_{\Lambda} P_{X|S\Lambda} P_{Y|ST\Lambda} P_{\Lambda}, 
\label{cslmodel}
\end{align}
for some choice of conditional probability distributions $P_{X|S\Lambda}, P_{Y|ST\Lambda},$ and $P_{\Lambda}$.

 A second class of causal accounts consists of those models that posit that 
 there is a latent variable that causally influences not only the two outcomes, but also one or both of the setting variables.  
We refer to such models as  parametrically {\em classical} and  structurally {\em superdeterministic}, abbreviated as \textsc{cSD}.
 (See Ref.~\cite{wood2015lesson} for a justification of this causal-modelling perspective  on
  the hypothesis of superdeterminism.)
We will again consider one particular representative from this class, 
corresponding to assuming the DAG depicted in Fig.~\ref{DAGs}(d). \rob{We denote this model by \textsc{cSD}$_0$, where the subscript simply serves as a reminder that there are other representatives.}
 Again, we consider each possibility for the cardinality of the latent variable $\Lambda$ as a distinct model, and we \rob{take the model to be parameter-unrestricted.}
Because the setting $S$ has a causal parent in this DAG, we must explicitly condition on $S$ to obtain the conditional $P_{XY|ST}$.
That is, 
$P^{{\rm \textsc{cSD}}_0}_{XY|ST} = P^{{\rm \textsc{cSD}}_0}_{XYS|Y} / P^{{\rm \textsc{cSD}}_0}_{S|T}$. 
Consequently, the compatible correlations in this case (for a fixed cardinality of $\Lambda$) are those that can be written as
 \begin{align}
P^{{\rm \textsc{cSD}}_0}_{XY|ST}
&:=  \frac{ \sum_{\Lambda} P_{X|S\Lambda} P_{Y|T\Lambda} P_{S|\Lambda} P_{\Lambda}}{\sum_{\Lambda'} P_{S|\Lambda'} P_{\Lambda'}}. 
\end{align}
for some choice of conditionals $P_{X|S\Lambda}$, $P_{Y|T\Lambda}$,  $P_{S|\Lambda}$ and $P_{\Lambda}$.
The final causal model we consider is one that is structurally of the common-cause form, just as \textsc{cCC} is, but  parametrically {\em quantum},  abbreviated \textsc{qCC} and
 depicted in Fig.~\ref{DAGs}(e).  Here, the latent common cause consists of the composite of the pair of quantum systems prepared in the Bell experiment, denoted $AB$.  The associated node in the DAG is depicted differently from the others  as a reminder that the parameters which make reference to it  are conditional density operators rather than conditional probability distributions.
The distributions over the observed variables that are deemed compatible are computed from an expression similar to Eq.~\eqref{clcmodel}, but where do-conditionals 
 are replaced by conditional density operators.  In the notation of Ref.~\cite{leifer2013towards}, this expression is
\begin{align}
P^{\;{\rm \scalebox{.6}{\hspace{-.15cm}\textsc{qCC}}}}_{XY|ST} &:= {\rm Tr}_{AB} \big( \rho_{X|SA} \;\rho_{Y|TB}\;  \rho_{AB} \big).\nonumber
\end{align}
\rob{for some choice of conditional density operators $\rho_{X|SA}$, $\rho_{Y|TB}$, and $\rho_{AB}$}.  
In a more conventional notation, the compatible distributions 
are those whose components can be expressed as:
\begin{align}
P^{_{\rm \scalebox{.6}{\hspace{-.001cm}\textsc{qCC}}}}_{XY|ST}(xy|st) &:= {\rm Tr}_{AB} \big[ (E^A_{x|s}\otimes E^B_{y|t}) \rho_{AB} \big],\label{qlcmodel}
\end{align}
\rob{for some choices of $\{ E^{A}_{x|s}\}_{x}$ and $\{ E^{B}_{y|t}\}_{y}$, which are positive operator-valued measure on systems $A$ and $B$ respectively, and for some choice of $\rho_{AB}$, which is a density operator on the bipartite system $AB$.}
 Note that one obtains a distinct model for every choice of Hilbert-space dimension for $A$ and $B$.


\textit{Model Selection.} Each causal model we consider defines, via Eqs.~(\ref{clcmodel})-(\ref{qlcmodel}), a set of correlations
 that are compatible with it. Each of these sets constitutes a statistical model. The problem of causal model selection,  therefore, reduces to statistical model selection. We denote the set of all possible distributions compatible with a causal model $\textsc{M}$ by $\mathcal{P}_{\textsc{M}}$. 
 


Consider the problem of identifying which values of the parameters in a causal model best fit a given set of observed data.  \rob{Quality of fit is measured by a loss function between the set of correlations realized by those parameter values (the realized statistical model) and the observed relative frequencies.} The best-fit model is the one that minimizes this loss function.  
  We will use the squared error loss function, which for our data corresponds to
 \begin{equation}
\rob{   {\rm loss}(P,F) =  \sum_{s,t,x,y}\big[ P_{XY|ST}(xy|st) -F(xy|st) \big]^2,   }
   \label{loss}
\end{equation}
 where $F(xy|st)$  is the observed relative frequency of outcomes $X=x$ and $Y=y$ given settings $S=s$ and $T=t$, while $P_{XY|ST}(xy|st)$ is the probability of outcomes $X=x$ and $Y=y$ given settings $S=s$ and $T=t$ for a particular choice of parameters in the model.  \rob{Here, $F$ and $P$ denote the matrices whose components are $F(xy|st)$ and $P_{XY|ST}(xy|st)$ respectively.}
We opt for this loss function as it is the most common choice\footnote{Although we also performed the data analysis using other loss functions and the conclusions remained the same.}.  \rob{The set of data that one uses to optimize the model parameters is called the {\bf training set}, denoted $F_{\rm train}$.}
The minimum value of the loss achieved by a model $M$ in a variation over parameter values consistent with that model is termed the {\bf training error} for $M$,
\begin{equation}
\rob{   {\rm TrainingErr}_{\textsc{M}} = \min_{P\in \mathcal{P}_{\textsc{M}}} 
{\rm loss}(P,F_{\rm train})   }
   \label{L2loss}
\end{equation}
The set of correlations defined by the best-fit model is consequently
 \begin{equation}\label{bestfitmodel}
\rob{\hat{P}^{\textsc{M}}= \underset{P\in \mathcal{P}_{\textsc{M}}}{\rm argmin}  \ {\rm loss}(P,F_{\rm train})   }
\end{equation}


Returning to the problem of model selection, the reason one cannot simply select the model with the smallest training error is that this would fail to take into account overfitting.
For example, in the case of a pair of models, $\textsc{M}$ and $\textsc{M}'$, 
where there is strict inclusion of the sets of compatible distributions, $\mathcal{P}_{\textsc{M}'} \subset \mathcal{P}_{\textsc{M}}$,
such a selection criterion would always prefer the model with the largest set of compatible distributions,
  even though the latter may be less predictively accurate. 
A more appropriate criterion is to select the model that minimizes the predictive error for independent data, a quantity termed 
 the {\bf test error}~\cite{hastie2003elements}. 

Strictly speaking, the test error of a statistical model is defined as its quality of fit with the true underlying distribution.
However, the latter is unknown, and so in practice one makes use of an estimate of the test error.  We follow a standard approach for data-rich problems, wherein \rob{one estimates the test error of one's statistical model using a second data set} called the {\bf test set} and denoted $F_{\rm test}$. 
 Specifically, we use what is termed the {\em plug-in estimate} of the test error~\cite{wasserman2013all}, 
\begin{equation}
     {\rm \widehat{TestErr}}_{\textsc{M}} = {\rm loss}( \hat{P}^{\textsc{M}}, F_{\rm test})
     \label{ExpectedTEloss},
\end{equation}
\rob{where  $\hat{P}^{\textsc{M}}$ is as defined in Eq.~\eqref{bestfitmodel}, namely, the statistical model defined by causal model $M$ and parameter values that yield the best fit to the training set. }
If the test set is large, this is likely to be a good estimate \rob{of the true test error of the statistical model $\hat{P}^{\textsc{M}}$}. 
Our criterion for model selection is minimization of the estimated test error.   
Hereafter, we will refer to the estimated test error as simply the test error.

Note that for the \textsc{cCE}$_0$ and \textsc{cSD}$_0$ models, we treat each possibility for the cardinality of $\Lambda$ as a separate model and find the one with the most predictive power.
Thus, the test error reported for these
 is the {\em minimal} value in a variation over this cardinality.  (We take a similar approach to cardinality in fits to the \textsc{cCC} causal model, to be discussed below.)
For the quantum model,  where the common cause is modelled as a bipartite quantum system $AB$ rather than a variable $\Lambda$,  the analogue of the cardinality of $\Lambda$ is the dimension of the Hilbert space describing $AB$. 
Because we found that the model wherein $AB$ is a pair of qubits 
already outperformed the other causal models on our slate of candidates, we did not explore \textsc{qCC} models with higher-dimensional common causes.


\begin{figure*}[t!]
\begin{center}
\includegraphics[scale=0.3]{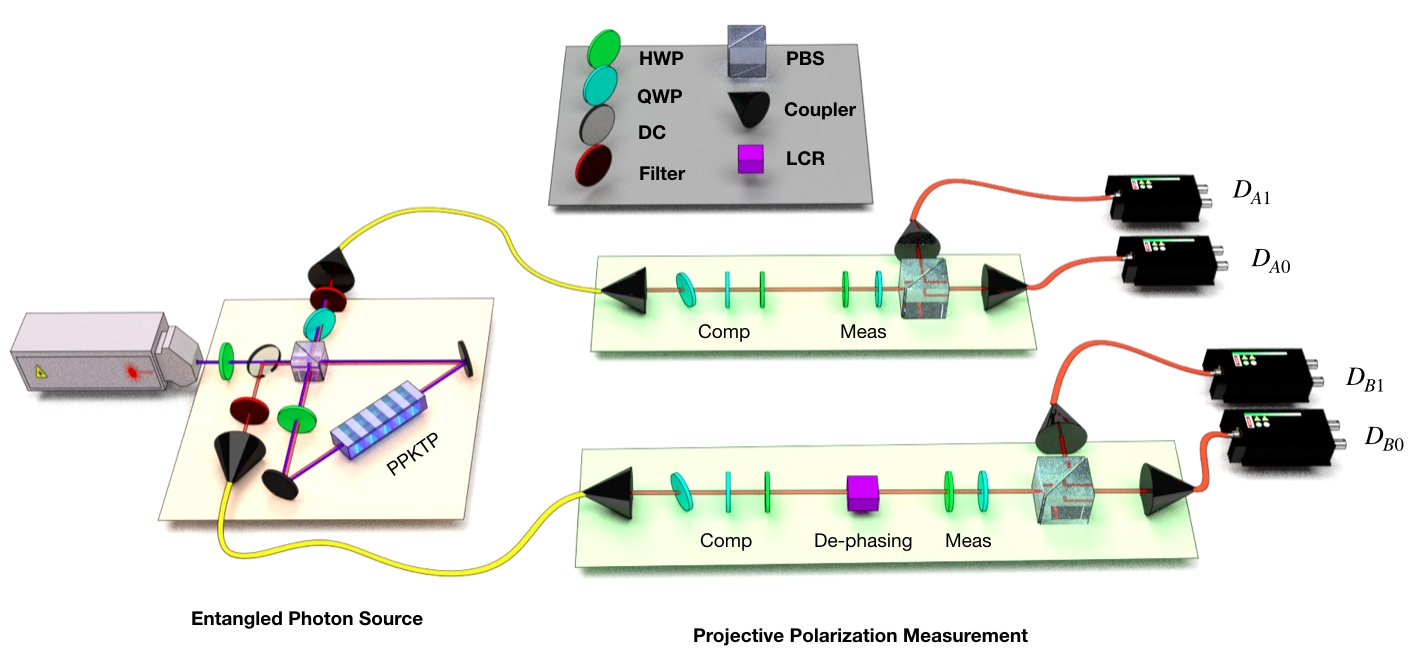}
\caption{Experimental diagram.  Maximally polarization entangled photons pairs are created through parametric down conversion in both paths of a Sagnac interferometer.  After compensating for the drift in the fiber-optic, each photon is sent to a polarization measurement, where the choice of measurement is controlled by half-wave plates, and coincidence counts 
are recorded.  In the dephased version of the experiment (discussed in Appendix~\ref{dephased}), 
a dephasing channel based on an LCR is implemented on one of the photons prior to measurement, while in the entangled version of the experiment, this channel is absent.
  PPKTP,  periodically-poled  potassium  titanyl  phosphate;  PBS,  polarizing beamsplitter;  LCR,  liquid  crystal  retarder;  HWP,  half-wave  plate;  QWP,  quarter-wave plate; DC, dichroic mirror}
\label{Experimental Diagram}
\end{center}
\vspace{-2em}
\end{figure*}

\textit{Results.} In our experiment, polarization entangled photons are generated using type-II spontaneous parametric down-conversion (Fig.~\ref{Experimental Diagram}) at a rate of 22000 singles/s and 800 coincidences/s with a 3ns coincidence window. The source produces the state $\frac{1}{\sqrt{2}}(\ket{HH}+\ket{VV})$ with $97.9\pm0.07\%$ fidelity. The photons are sent to different polarization analyzers, functioning as Alice's and Bob's labs, each one implementing one of six possible binary-outcome measurements. The data was collected for 10$s$ for each pair of values $(s,t)$ of the measurement setting variables,
and the relative frequency $F(xy|st)$ with which Alice obtains outcome $x$ in coincidence with Bob obtaining outcome $y$ was recorded. \rob{The Poissonian noise model for the photon counts is used to generate bootstrap estimates of the confidence intervals.} 
The entire experiment was performed twice, thereby yielding a training data set and a test data set. 
These two data sets are the only inputs to our causal discovery algorithm.

\begin{figure}[htb]
\begin{center}
\includegraphics[scale=0.24]{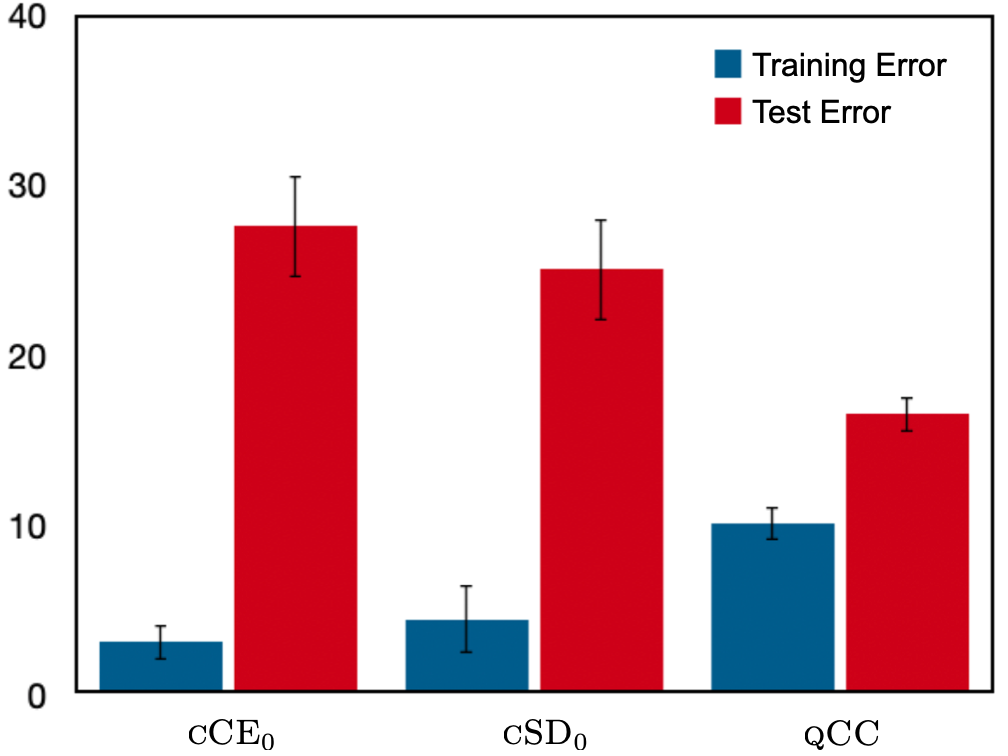}
\caption{
Adjudicating between different causal models based on the experimental data.
 Plotted are the training error (blue) and test error (red) for the \textsc{cCE}$_0$, \textsc{cSD}$_0$ and \textsc{qCC} models. 
Error bars denote a confidence region of one standard deviation. The \textsc{qCC} model has the lowest test error and is therefore preferred. The fact that the larger test error of the \textsc{cCE}$_0$ and \textsc{cSD}$_0$ models is accompanied by {\em a lower training error}
suggests that they {\em overfit} the data relative to \textsc{qCC}.}
\label{EntangledLoss}
\end{center}
\vspace{-2em}
\end{figure}

Given that our purpose here is to adjudicate between models that {\em can} account for a violation of Bell inequalities, we leave aside $\textsc{cCC}$, and focus on adjudicating between  $\textsc{cCE}_0$, $\textsc{cSD}_0$, and $\textsc{qCC}$. 
  The training errors and the test errors for each are shown in Fig.~\ref{EntangledLoss}. Recall that the selection criterion is minimization of the test error.
As the difference in test error between  \textsc{cCE}$_0$ and \textsc{qCC} is approximately 5 standard deviations,
 and there is a similar gap between  \textsc{cSD}$_0$ and \textsc{qCC}, it follows that \textsc{qCC}  emerges as the preferred model with high statistical confidence.   

\textit{Discussion.} We now address the question of  {\em why} \textsc{qCC} tests better (i.e., achieves a lower test error) than \textsc{cCE}$_0$ and \textsc{cSD}$_0$.  The fact that both \textsc{cCE}$_0$ and \textsc{cSD}$_0$ {\em train} better (i.e., achieve a lower training error) than \textsc{qCC}
  provides some insight into why this is the case.  When a model trains better but tests worse than another, a likely explanation is that the first is more prone to overfitting---achieving a better fit to the training data by fitting to statistical fluctuations found therein---and this in turn implies a worse fit to the test data. It is likely, therefore, that \textsc{cCE}$_0$ and \textsc{cSD}$_0$ overfit the data as compared
 to \textsc{qCC}.

 In a Bell experiment, there is statistical {\em independence} between an outcome variable at one wing and the setting variable at the opposite wing.  This is typically termed the `no-signalling condition'.  Classical causal models that are structurally radical and \rob{parameter-unrestricted}
  can reproduce the no-signalling condition, but, as shown in Ref.~\cite{wood2015lesson} (see also Ref.~\cite{cavalcanti2018classical}), they can only do so for a {\em special class} of values of the parameters \rob{(which is a set of measure zero in the full set of possible parameter values, so that they {\em require fine-tuning} in order to do so).}
    Given that any finite sample of data exhibits statistical fluctuations {\em away} from such independence, in a structurally radical classical causal model, it is possible for the fitting procedure to mistake these fluctuations for real features, thereby yielding best-fit values of the parameters outside of the special class.  In short, such models have an opportunity to overfit the data.  
For structurally conservative models like \textsc{qCC}, on the other hand,  there is no possibility of such overfitting because  the no-signalling condition is implied by the causal structure and therefore
  holds for {\em all} choices of the parameter values.  
This is the reason, we believe, that \textsc{cCE}$_0$ and \textsc{cSD}$_0$ overfit the data as compared to \textsc{qCC}.

Although we have here considered only the \textsc{cCE}$_0$ representative of the \textsc{cCE} class of models and only the \textsc{cSD}$_0$ representative of the \textsc{cSD} class, similar considerations apply for other representatives.  More precisely, for {\em every} DAG in these classes---regardless of what pattern of interlab influences or superdeterministic common causes they posit---as long as the model is parameter-unrestricted,
 the no-signalling condition is only reproduced for a special class of parameter values~\cite{wood2015lesson}.  Consequently, all such models are likely to be found to have less predictive power than $\textsc{qCC}$ by virtue of overfitting. 

Note that, relative to this account of the overfitting, one expects to obtain similar results 
 even if the quantum source is dephased in such a way that  the bipartite state it prepares is unentangled.
  We confirmed this expectation by performing a dephased version of our experiment and verifying that although the \textsc{cCC} model now performs comparably to the \textsc{qCC} model (since it also satisfies the no-signalling condition for all parameter values), the \textsc{cCE}$_0$ model still trains noticeably better and tests noticeably worse than either the \textsc{cCC} or \textsc{qCC} models, while the  \textsc{cSD}$_0$ model trains marginally better and tests marginally worse.  This lends further support to our interpretation of the overfitting.  Details are provided in Appendix~\ref{dephased}.

\textit{Conclusions.} 
In this article, we have confined our attention to causal model types wherein no restriction is imposed on the possible values of the parameters.
The fact that \rob{the classical causal models that}
  posit 
 inter-lab causal influences are disfavoured relative to models of the \textsc{qCC} type
  implies that one does not need to seal the locality loophole in a Bell experiment (i.e., perform measurements at space-like separation) in order for it to provide evidence of quantumness.
   In addition, the fact that the classical causal models that are superdeterministic are disfavoured relative to models of the \textsc{qCC} type overturns the claim that the loophole associated to the possibility of superdeterministic models
    cannot be closed. 

\rob{We now consider what conclusions can be drawn from our results and our data analysis technique if one relaxes the assumption that there is no restriction on parameter values, that is, what conclusions one can draw for {\em parameter-restricted} causal models.}


If  the range of parameter values is restricted in such a way that the model is compatible with {\em all and only}
 the correlations achievable in operational quantum theory,
 so that in particular the no-signalling condition is satisfied for all parameter values,
  then one cannot hope to experimentally distinguish it from \textsc{qCC} via our model selection technique.  The standard view of Bohmian mechanics~\cite{sep-qm-bohm} is likely to be an example of such a model (which is structurally radical  by virtue of allowing inter-lab causal influences).  Presumably, one can also construct superdeterministic   models of the Bell experiment that are of this type. 

If, on the other hand, the range of parameter values is restricted, but not such that it allows
{\em  only} those correlations achievable in operational quantum theory, then the model remains an {\em empirical competitor} to operational quantum theory.
 Many proponents of structurally radical ways out of the Bell no-go result \rob{do, in fact, endorse this type of model}
  and indeed take its
   empirical inequivalence to operational quantum theory  to be one of its
    virtues.  Valentini's subquantum-nonequilibrium version of Bohmian mechanics~\cite{valentini1991signal1,valentini1991signal2,valentini2002signal,valentini2002signaldeterministic}, which explicitly allows for the possibility of violations of the no-signalling condition in Bell experiments, is an example of such a model.  Because a model with a more restricted scope of parameter values can in principle exhibit {\em less} overfitting than its parameter-unrestricted counterpart, the conclusions of our analysis need not apply to these.  Nonetheless, these models can be included in the slate to which one applies the model selection technique described here.  It is merely a question of specifying the range of parameter values and restricting the optimization to this range.  A more detailed description of these conclusions is included  in Appendix~\ref{detailsconclusions}.

\rob{The results reported here serve as a constraint on }
the development of alternatives to quantum theory and of interpretations of the formalism.
More generally, \rob{the techniques we introduce have broad applicability in quantum foundations, as not just Bell-type experiments but {\em all} experiments seeking to adjudicate between different accounts of quantum phenomena} 
stand to benefit from a consideration of whether a given account thereof  {\em overfits} the data, rather than merely a demonstration that it does not underfit the data.

\acknowledgements

This research was supported in part by the Natural Sciences and Engineering Research
Council of Canada (NSERC), Canada Research Chairs,
Industry Canada, the Canada Foundation for Innovation (CFI), and the Canada First Research Excellence Fund (CFREF).
This research was also supported by Perimeter Institute for Theoretical Physics. Research at Perimeter Institute is supported by the Government of Canada through the Department of Innovation, Science and Economic Development Canada and by the Province of Ontario through the Ministry of Research, Innovation and Science. PD would like to thank Sacha Schwarz and Jean-Philippe MacLean for helpful discussions and tips.

\appendix

\section{Further details on the slate of causal models}\label{DetailsOnSlate}

In the framework for causal modelling that we are presuming, the modifiers `classical' and `quantum' refer only to the {\em parametric} part of a causal model.  The latter is considered classical if the parameters can be specified as conditional probability distributions, while it is considered quantum if the parameters are specified as conditional density operators (or, equivalently, completely positive trace-perserving maps). The structural part of a causal model is presumed to always be stipulated by a DAG.\footnote{The notion of a superposition of causal structures, advocated in some works~\cite{hardy2005probability,oreshkov2012quantum,chiribella2013quantum}, may resist formulation in terms of DAGs.  If so, then this possibility is excluded from the causal modelling framework we adopt here.  Note, however, that this limitation would  not be consequential for the purposes of this article, as we are not aware of any attempts to provide a causal account of Bell inequality violations which appeals to superpositions of causal structures.} 
If the causal structure mirrors the structure of the usual quantum expression for the correlations 
 when the latter is conceptualized as a contraction of tensors\footnote{See Ref.~\cite{hardy2012operator} for an articulation of this notion of structure for computations.}, then it is deemed conservative.  Otherwise, it is deemed radical.  \rob{In the case of a Bell experiment, the usual quantum expression for the correlations is given in Eq.~\eqref{qlcmodel} of the main text, so that an account has a conservative causal structure only if the DAG has the common-cause form of Figs.~\eqref{DAGs}(b) or (e).  }

 
 A quantum causal model of the experiment is defined in such a way that it can provide a causal account of the quantum predictions using the conservative causal structure, but where the price for the structural conservatism  is that the model must be  parametrically radical.  A classical causal model  of the same experiment, by contrast, salvages parametric conservatism \rob{by using only conditional probability distributions, but is thereby forced to be radical at the level of the causal structure.}




The causal-modelling perspective on the different ways out of the Bell no-go result was initiated in Ref.~\cite{wood2015lesson}.  That article stipulated how to recast within the framework of classical causal models the various traditional causal accounts of Bell inequality violations, all of which 
are parametrically conservative (i.e., parametrically {\em classical}) and structurally radical.  These are the accounts that invoke interlab cause-effect relations or superdeterminism.  We say a bit more about each case here.


The \textsc{cCE} class of causal models of the Bell experiment, wherein one posits a cause-effect influence between the labs, subsumes several possibilities for the causal structure.  


In this article, we focussed on one such structure, 
wherein there is a causal influence from the setting $S$  to the outcome $Y$.  By symmetry, of course we could  just as well have considered the model wherein there is a causal influence from the setting $T$  to the outcome $X$.  A distinct model within the \textsc{cCE} class is one wherein there are setting-to-outcome causal influences in {\em both} directions, that is, $S\rightarrow Y$ {\em and} from $T \rightarrow X$.
 Another such model is one wherein there is a causal influence from the {\em outcome} rather than setting on one wing to the outcome on the other wing, such as $X \rightarrow Y$. 
 More generally, every combination of influences between $S, X$ and $T, Y$ (e.g., every combination of arrows $X \rightarrow Y$, $S \rightarrow Y$, $S\rightarrow T$, $X\rightarrow T$,  $Y \rightarrow X$, $T \rightarrow X$, $T\rightarrow S$, $Y\rightarrow S$) that does not introduce cycles when added to the DAG of Fig.~\ref{DAGs}(b) constitutes an element of the \textsc{cCE} class of causal models. \rob{For all such models, the no-signalling condition is not implied by  the causal structure. }
   As we noted in the main text, the results of Ref.~\cite{wood2015lesson} \rob{imply that---assuming we are correct in our assessment that models that can violate the no-signalling condition tend to overfit the data---} {\em all} models \rob {in this class that can realize the quantum correlations}
    will be found to overfit the data and therefore to test poorly. 

It is important to note that there are models in this class for which the modifications to the causal structure, radical as they are, still {\em underfit} quantumly realizable data and consequently do not even {\em train} well on such data.
An example is the classical causal model wherein the {\em outcome} at one wing (rather than the setting) has a causal influence on the outcome on the other wing, for instance, one where the DAG of Fig.~\ref{DAGs}(b) in the main text is supplemented by an arrow from $X$ to $Y$. 
 For the case of ternary setting variables and binary outcomes, such a causal model was shown in Ref.~\cite{chaves2015unifying} to imply Bell-like inequalities that are quantumly violated.  As such, one expects that the performance of such models on experimental data that violates such inequalities would be comparable to the performance of the \textsc{cCC} model on our experimental data (see Sec.~\ref{cCCmodel}). 
That is, insofar as such models would {\em underfit} the data, they would both train and test poorly. \rob{The same comments apply to  causal structures wherein the outcome at one wing has an influence on the setting at the other wing.}
We noted in the main text that structurally radical models have the capacity to describe violations of the no-signalling condition, which ultimately is what causes them to overfit the data relative to the structurally conservative model.  An example helps to illustrate this point. Within the context of a pilot-wave theory, Valentini has proposed~\cite{valentini1991signal1,valentini1991signal2,valentini2002signal,valentini2002signaldeterministic} that obtaining the predictions of operational quantum theory might be contingent on the hidden variables being in a state of subquantum equilibrium, and that 
states of subquantum {\em non}equilibrium---allowing for deviations from the predictions of operational quantum theory---might be possible in exotic scenarios.  In particular, 
 the proposal explicitly allows for 
violations of the no-signalling condition in subquantum nonequilibrium.  



We now turn to the \textsc{cSD} class of causal models of the Bell experiment: classical causal models that posit that one or both of the setting variables share a common cause with the outcome variable \rob{at the opposite wing}.  \rob{If one considers the case where the setting $S$ shares a common cause with the outcome $Y$, then the resulting causal model is subsumed within the causal model for which the causal structure includes a three-way common cause of $S$, $X$ and $Y$.  The latter is the one depicted in Fig.~\ref{DAGs}(d) and which we have focussed on in this article.}
\rob{Note that the superdeterminist view is sometimes described as positing a common cause $\Lambda'$ between a setting variable (say $S$) and the hidden variable $\Lambda$ in the DAG of Fig.~\ref{DAGs}(b), but without loss of generality one can absorb $\Lambda$ into the definition on $\Lambda'$, thereby obtaining the causal structure that posits a common cause of $S$, $X$ and $Y$.  
}

\rob{To obtain a distinct representative of the \textsc{cSD} class, it suffices to swap the roles of the pair of parties relative to the DAG of Fig.~\ref{DAGs}(d) and 
have $\Lambda$ be a common cause of $T$, $X$ and $Y$.
  Alternatively, one could presume that $\Lambda$ influences not only the pair of outcome variables, but {\em both} setting variables as well.  }

\color{black}


As noted in Ref.~\cite{wood2015lesson}, generic values of the parameters in such models also yield nontrivial correlations between an outcome variable and the setting variable in the opposite lab, hence a violation of the no-signalling condition.  As in the \textsc{cCE} case, it is this possibility that makes such models prone to 
overfitting the data relative to the structurally conservative model.

The distributions compatible with a given classical causal model depend on the cardinalities of the latent variables.
Suppose a given causal model incorporates a latent variable $\Lambda$. The set of distributions that are compatible with the classical causal model where $\Lambda$ has cardinality $d$  contains the set of distributions where $\Lambda$ has cardinality $d'$ for any $d'<d$.
 Even if one is considering the true causal structure, therefore, if one assumes too low a cardinality for the latent variables, one might still underfit the data.  If not underfitting the data were the {\em only} standard of empirical success,  it would be reasonable to presume the cardinalities of the latent variables to be those that are just sufficient to saturate the set of distributions compatible with the causal structure (these are known to be finite for classical causal models where the observed variables are discrete~\cite{rosset2018universal}).   This choice, however, may {\em overfit} the data relative to a model with smaller cardinalities.  
  This is why, for a given causal structure, we treat each possibility for latent cardinalities as a separate representative of its class of causal models, a separate model in the slate of candidates.


The \textsc{qCC} model posits the conservative causal structure, that is, a common cause acting between the labs, but it requires the parameters to be conditional density operators rather than conditional probability distributions, as noted in the main text.  

For such a model, the common cause is modelled as a bipartite quantum system $AB$ rather than a variable $\Lambda$,  so the analogue of the cardinality of $\Lambda$ is the dimension of the Hilbert space describing $AB$. Nonetheless, the same considerations hold: a dimension that is higher than the optimum may lead to overfitting of the data while a dimension that is lower than the optimum may lead to underfitting of the data, so it is best to treat each possibility for the dimension as a separate representative of the class of quantum causal models. As noted in the main text, however, the model wherein $A$ and $B$ are both qubits, so that $AB$ has dimension 4, already outperformed the other causal models on our slate of candidates, and so we did not need to consider any variation in this dimensionality.




\subsection{On whether the \textsc{qCC} model provides a satisfactory causal explanation}

The idea that the best causal account of Bell inequality violations is achieved by \textsc{qCC}, that is, a causal model with a quantum common cause, is one that has been espoused in Refs.~\cite{leifer2013towards, wood2015lesson, allen2017quantum,wolfe2020quantifying}.  
As was acknowledged in those articles, however, 
whether 
such a causal model can really be deemed to achieve a {\em realist} account of the experimental data (as structurally radical classical causal models do) is one that has not yet been adequately answered.  Ref.~\cite{schmid2020unscrambling} argues that providing an affirmative answer to the question  
depends on the success of a research program that seeks to modify the notions of causation and inference to achieve a novel, nonclassical, type of realism that can underlie operational quantum theory.\footnote{In terms of the framework introduced in Ref.~\cite{schmid2020unscrambling}, the present article is about adjudicating between different ``quotiented operational causal-inferential theories'' based on their power to predict the experimental data.  It is not directly about adjudicating between realist representations thereof.  However, the assumption that any realist representation of an operational causal-inferential theory 
 must be {\em diagram-preserving} means that the latter must posit the same causal structure as the former for any given experiment.  (See Section V.A of Ref.~\cite{schmid2020unscrambling} for a defense of the assumption of diagram preservation.)   Consequently, if one provides experimental evidence against a given causal structure for operational accounts of the experiment, one has provided experimental evidence against that causal structure for realist accounts as well. }

\section{Further details on the causal discovery algorithm}
 


We here provide further information about the algorithm we propose for adjudicating between causal models.
 A {\em classical} causal discovery algorithm~\cite{pearl2000causality,Spirtes00} adjudicates between different classical causal models based on purely observational data. 
The algorithm we describe is also based on purely observational data, but 
   the slate of candidate causal models is allowed to include {\em intrinsically quantum} causal models and allows a direct comparison between these and classical causal models using the same scoring criteria.

In the statistics community, in order to accommodate data-poor applications, a variety of techniques have been developed to estimate the test error while still using {\em all} of the acquired data for fitting the model parameters.  Common techniques of this sort include the Akaike Information Criteria (AIC), the Bayesian Information Criteria (BIC) and Cross Validation (CV).    The experiment we consider, however, is data-rich, and consequently we can simply fit the model parameters using one data set, termed the {\em training set}, and estimate the test error using a second data set, termed the {\em test set}.  

\begin{figure*}[t!]
\begin{center}
\includegraphics[scale=0.36]{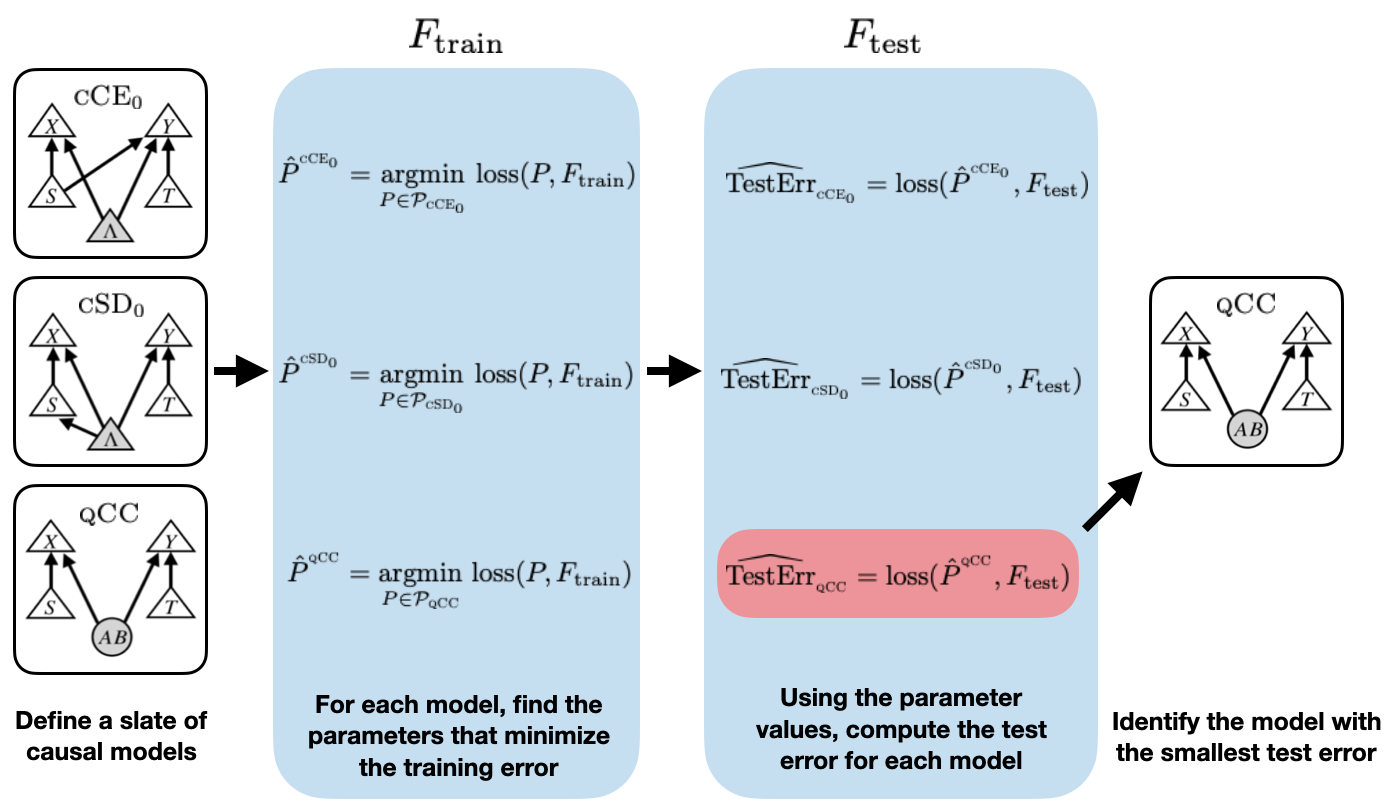}
\caption{A schematic of our model selection technique, where the quantities referenced are defined in Eqs.~\eqref{loss}, \eqref{bestfitmodel}, and \eqref{ExpectedTEloss} of the main text. 
}
\label{algorithm}
\end{center}
\vspace{-2em}
\end{figure*}

A schematic of the algorithm is provided in Fig.~\ref{algorithm}. The input of the algorithm is the observed relative frequency $F(xy|st)$ of outcomes $X=x$ and $Y=y$ given settings $S=s$ and $T=t$ for the training set and the test set.
The output of the algorithm is the training error and the estimated test error, as well as their standard deviations, for each of the causal models on the slate of candidates.  These allow for the determination of an ordering of the models by their relative predictive power, as well as an estimate of the statistical confidence of this ordering

For each causal model $M$, one minimizes the training error over the conditional probability distributions in the set $\mathcal{P}_M$ that are compatible with the model. 
 As noted in the main text, we take the least squared error as our loss function.  We note, however, that the data analysis was repeated for three other loss functions and the conclusions were essentially unchanged. 
 

The set of distributions that are compatible with a generic causal model $M$, that is, $\mathcal{P}_{M}$, is generally a nonconvex set~\cite{wolfe2019inflation}.  In particular, this is the case for the causal models of interest to us here whenever the cardinality of $\Lambda$ or the dimension of the $AB$ system are not maximal. Consequently, our technique requires a nonconvex optimization, which is generally difficult as there can be multiple local minima.
Indeed, the optimization comprises almost the entire computational burden of the algorithm. 
To do so, we use the Nelder-Mead method, implemented via NMinimize in Mathematica.
We repeated each optimization 100 times which different random seeds in order to increase the chances of finding the global minimum. 
Given that our loss function is an {\em estimate} of the test error, identifying a local minimum rather than the global minimum provides a
 worse estimate of the true test error. However, there is no reason to think that the model is more likely to get stuck in any one local minimum rather than another. 


The next step is to calculate, for each model on the slate of candidates,  the estimated test error for the distribution that was found to minimize the training error. 
 This is a relatively 
  inexpensive computation.  The model that is favoured by the model selection technique is the one which is estimated to have the most predictive power, that is, the one that achieves the smallest estimated test error.
 
 Error bars for the training and test error are calculated by doing parametric bootstrap re-sampling 
  of the initial frequency counts in the data sets (a type of Monte Carlo error estimation). The entire algorithm is repeated to find the test and training errors for the re-sampled counts. We performed 10 re-samplings and used the empirical standard deviation of this sample as our error bars.

As was noted in Section~\ref{DetailsOnSlate}, in order to find the most predictive model with a given causal structure,  one must allow for variability in the cardinality of the latent variables therein.  A brute-force optimization over the choice of these cardinalities, however, is computationally inefficient.  To reduce computation time, therefore, we introduce a heuristic for this optimization, which we now describe for the case of the Bell experiment.  For a given structurally radical hypothesis, such as \textsc{cCE}$_0$ or \textsc{cSD}$_0$, we order the causal models associated to it by the cardinality of $\Lambda$.  We then search through increasing values of cardinality for the most predictive model, and we end the search if either of the following two conditions are met: (i) for three consecutive values of the cardinality,  the training error decreases while the estimated test error increases, or (ii) for three consecutive values of the cardinality,  the training error and estimated test error take the same value.  The first condition is reasonable since it is likely to describe a situation wherein the increasing expressive power of the models is yielding increasing degrees of overfitting, so that continuing to increase the expressive power by increasing the cardinality would only further decrease the predictive power.  The second condition is reasonable since it is likely to describe a situation wherein the increasing expressive power of the models is not yielding increasing degrees of overfitting, but also not improving the fit, so that increasing the cardinality will not lead to any further increases in predictive power.\footnote{This can occur, for instance, if the cardinality has increased beyond the value that saturates the set of compatible distributions, supposing one does not know a priori what this value is. }




\subsection{Comparison to prior work}\label{priorwork}

As noted above, the fact that our model selection technique is applicable to purely observational data makes it a generalization of
the standard notion of a {\em causal discovery algorithm} in the causal inference literature~\cite{pearl2000causality,Spirtes00}.
 Ref.~\cite{ried2015quantum} also considered a quantum generalization of the notion of causal discovery, 
but only for the special case of distinguishing a cause-effect relation from a common-cause relation and only device-dependently.  
Several other works~\cite{CostaShrapnel,giarmatzi2019quantum,bai2020efficient} have focussed on the problem of determining the causal structure based on {\em interventionist} data.\footnote{These works have also described their algorithms as instances of `causal discovery', although the usage of this term is somewhat at odds with the convention in the causal inference community of reserving this term for analyses based on observational data.}   That is, rather than making inferences about causal structure based on a probability distribution over the observed classical variables, as we do here, they do so based on a tomographic characterization of each process in a circuit that describes the causal relations.

\section{Further details on the experiment}

\subsection{Selection of the set of measurements}

\begin{figure}[h]
\begin{center}
\includegraphics[scale=0.45]{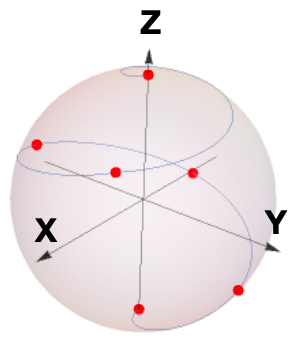}
\caption{A qubit Bloch sphere. Each of the six red points correspond to the zero outcome for one of the six binary projective measurements each lab attempted to implement. These measurements were chosen so that they are distributed roughly uniformly around the surface of the sphere.}
\label{bloch_sphere}
\end{center}
\vspace{-2em}
\end{figure}

In this section, we describe the motivation behind the choice and number of measurements we implemented in our Bell experiment. We sought to perform the minimal number of measurements which could distinguish the various causal models with high statistical confidence.
Minimizing the number of measurements is desirable because it limits the time required to complete the experiment and therefore also limits the error introduced by temporal drift of the experimental components.
 We did not attempt, however, to find provably optimal sets. 
 Instead, we considered sets of binary-outcome projective measurements chosen in such a way as to be spread out over the Bloch sphere.  For different values of $n$, we chose $n$ rank-1 projectors that were equally spaced along a 1-parameter family that traces a spiralling curve on the Bloch sphere (see Fig.~\ref{bloch_sphere}). 
We then determined, by simulating data for an idealized version of our experiment,
 the minimal value of $n$ for which our causal discovery algorithm could distinguish the slate of causal models under investigation.  In this way, we settled on implementing {\em six} binary-outcome measurements at each side of our Bell experiment. 


Our simulations indicated, in particular, that if instead of the six measurements along the Bloch sphere spiral, one were to implement the standard pair of measurements appearing in the Clauser-Horne-Shimony-Holt (CHSH) version of the Bell experiment, then the \textsc{cCE}$_0$, \textsc{cSD}$_0$ and \textsc{qCC} models had scores within one standard deviation of each other (for a quantity of data that was matched to what we could hope to achieve experimentally), such that our model selection technique could not reliably adjudicate between them.  
  By contrast, for the six binary-outcome measurements described in Fig.~\ref{bloch_sphere}, the test error of the most predictive model (\textsc{qCC}) was separated from that of the other models (\textsc{cCE}$_0$ and \textsc{cSD}$_0$) by approximately 10 standard deviations \rob{in our simulations, thereby suggesting that this choice would be adequate for our purposes (which the experiment and our data analysis subsequently confirmed).}

It is at present not clear how to anticipate by theoretical considerations rather than numerical simulations
  the minimum number of measurements required by our model selection technique to achieve a given level of statistical confidence.

\subsection{Performance of the model that is parametrically classical and structurally common-cause}\label{cCCmodel}

We noted in the main text that the \textsc{cCC} model, by virtue of being incapable of violating the Bell inequalities, cannot possibly do justice to our experimental data.  For this reason, it was not included it in the slate of candidate causal models among which we implemented model selection.  Nonetheless, confirming the expectation that the \textsc{cCC} model is not viable as a causal account of our Bell experiment provides an additional check on our model selection technique, and so we do so here.  Note that our technique does not establish the nonviability of this model in the standard way, that is, by demontrating that 
our experimental data violates a Bell inequality by many standard deviations.  Rather, it considers the training error and the test error of the \textsc{cCC} model for our data.  
 The model is found to have a training error of $790 \pm 10$,  far in excess of the training error of \textsc{cCE}$_0$, \textsc{cSD}$_0$ or \textsc{qCC} \rob{(reported in Fig.~\ref{EntangledLoss} of the main text)}, indicating that it underfits the data relative to them, as one would expect
  given its inability to violate Bell inequalities.  It consequently also has poor predictive power, achieving a test error of $800 \pm 10$, also far in excess of \textsc{cCE}$_0$, \textsc{cSD}$_0$ or \textsc{qCC}.  

\subsection{The dephased version of the experiment}\label{dephased}

We now describe the results of applying our model selection technique to the version of our experiment wherein the initial entangled state is completely dephased, so that it becomes a separable state. 

  Our primary motivation for doing so is to provide additional evidence for our interpretation of the results of the entangled version of the experiment.  As noted in the main text, the fact that \textsc{cCE}$_0$, \textsc{cSD}$_0$ train better but test worse than \textsc{qCC} suggests that they {\em overfit} the data relative to \textsc{qCC}. Specifically, the statistical fluctuations away from the no-signalling condition (which are possible in any finite-run data) can be mistaken as real features by \textsc{cCE}$_0$ and \textsc{cSD}$_0$, but not by \textsc{qCC}, since \textsc{qCC} satisfies the no-signalling condition for all parameter values.  But if this is indeed the case, then one expects analogous results in a dephased version of the experiment (which removes entanglement, leaving only classical correlations), where now the \textsc{cCC} model can do justice to the observations while also satisfying the no-signalling condition for all parameter values.  Because the dephased version of the experiment is also likely to exhibit statistical fluctuations away from the no-signalling condition, one expects that \textsc{cCE}$_0$ and \textsc{cSD}$_0$ can once again mistake these fluctuations for real features, while neither \textsc{qCC} nor \textsc{cCC} can do so, thereby leading \textsc{cCE}$_0$ and \textsc{cSD}$_0$ to overfit the data relative to \textsc{qCC} and \textsc{cCC}.



The dephased version of our experiment also provides another opportunity to check our model selection technique: given that it prepares a separable state and therefore cannot violate any Bell inequalities, one expects that the \textsc{cCC} model should have as much predictive power as the \textsc{qCC} model for this experiment.  
  
The entanglement between the two photons  is removed by implementing a completely dephasing channel in the basis of eigenstates of the Pauli $X$ operator (i.e., a noisy bit-flip channel) on one of the photons, via rapid switching between a Pauli $X$ gate and an identity gate. This is achieved using a liquid crystal retarder (LCR) that is switched 
 every 500$ms$. 
 The bipartite state after the dephasing channel is found to have $98.3\pm0.07\%$ fidelity with the equal mixture of $\frac{1}{\sqrt{2}}(\ket{HH}+\ket{VV})$ and $\frac{1}{\sqrt{2}}(\ket{HV}+\ket{VH})$, which is a separable state, namely, $ \frac{1}{2} |DD\rangle \langle DD| + \frac{1}{2} |AA\rangle \langle AA|$.
The rest of the experiment proceeds as before, as does the data analysis.
 
\begin{figure}[htb]
\begin{center}
\includegraphics[scale=0.24]{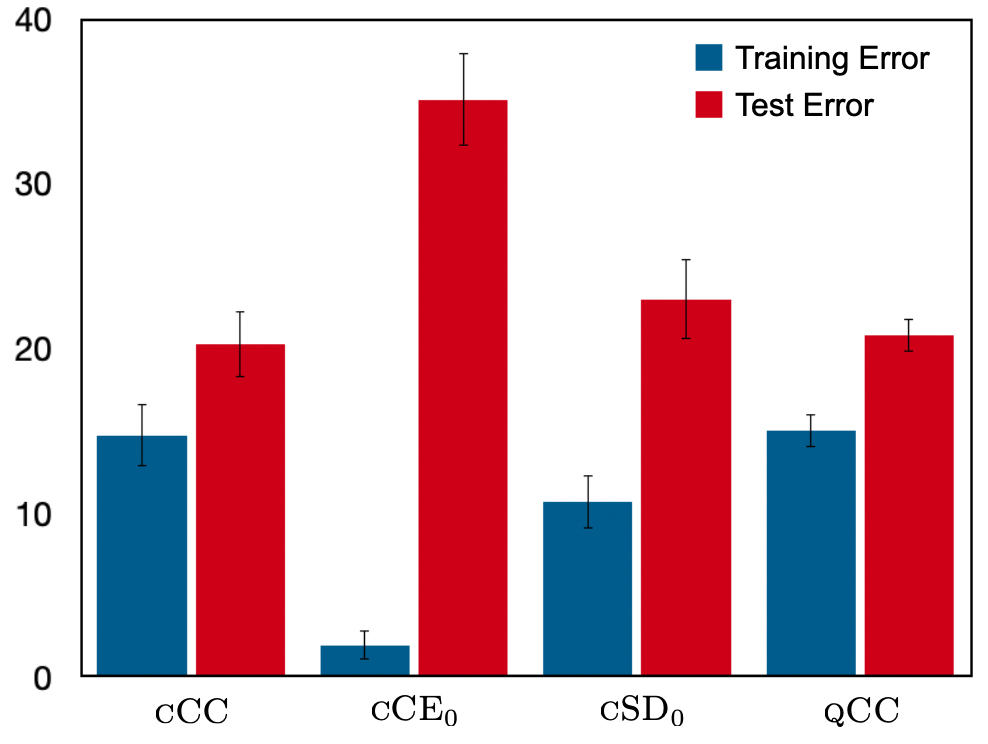}
\caption{The results of the experiment with the de-phased source.
Plotted are the training error (blue) and test error (red) for the \textsc{cCC}, \textsc{cCE}$_0$, \textsc{cSD}$_0$ and \textsc{qCC} models. 
 The error bars denote a confidence region of one standard deviation. 
 }
\label{DephasedLoss}
\end{center}
\vspace{-2em}
\end{figure}

 
 The results of the experiment are presented in Fig.~\ref{DephasedLoss}.  
We confirm that the \textsc{cCC} model has a training error and test error that are comparable (i.e., within error) to those of the \textsc{qCC} model.
We also see that the \textsc{cCE}$_0$ model trains noticeably better and tests noticeably worse than either \textsc{cCC} or \textsc{qCC},
  while the  \textsc{cSD}$_0$ model  trains marginally better and tests marginally worse than these, thereby confirming our expectations about their relative performance. 
  

\section{Further details on the conclusions that can be drawn from our results}\label{detailsconclusions}

\subsection{Some distinctions among causal models}

We wish to  consider in more detail what our results imply for various loopholes in Bell experiments.  To prepare the ground for this discussion, it is useful 
  to introduce some distinctions among causal models based on the scope of parameter values that they allow. We do so in a manner
that is not specific to the Bell experiment, so that such distinctions can be articulated for experiments with arbitrary compositional structures. 

%
 


Let the DAG corresponding to the conservative causal structure for an experiment be denoted by $G$, while structurally radical alternatives to it are denoted $G_0, G_1, \dots$.  Let the model of the experiment that is structurally conservative but parametrically quantum (i.e., parametrically radical)
   be denoted \textsc{q}$G$, and let the class of models that are 
 parametrically classical  (i.e., parametrically conservative) but structurally radical---in the sense of assuming a DAG $G_i$ that is not the conservative one---be denoted by  \textsc{c}$G_i$.
 
As noted in the main text, causal models can also be distinguished by what they assume about the cardinality of the latent variables in the DAG.    However, we shall not introduce additional notation for this distinction.  Rather we will presume that when a given class of causal models is considered as a candidate account of some data, it is the particular element of the class that {\em optimizes} the cardinality of the latent variables that is considered.
 

Within a given class \textsc{c}$G_i$, one can distinguish different models based on the fact that one can imagine restrictions on the parameter values allowed in the model.  There can be different types of restrictions on the parameters, each of which leads to a different model within the class. 

First, consider what it means to assume no such restriction.  For each variable $X$ corresponding to a node of $G_i$, the usual notion of a classical causal model allows {\em any} conditional probability distribution $P_{X|{\rm Pa}(X)}$ in the set $\mathcal{P}_{X|{\rm Pa}(X)}$ of all such conditionals.  As noted in the main text, we refer to such a classical causal model as the {\em parameter-unrestricted}  model for $G_i$. In the context of the Bell experiment,  \textsc{cCE}$_0$ is the parameter-unrestricted model for the DAG of Fig.~\ref{DAGs}(c), while \textsc{cSD}$_0$ is the parameter-unrestricted model for the DAG of Fig.~\ref{DAGs}(d).

But one can also define {\em parameter-restricted} classical causal models.  For one of more variables $X$ corresponding to the nodes of $G_i$, the possibilities for  $P_{X|{\rm Pa}(X)}$ are presumed to be restricted to   a set $\mathcal{P}^{\rm sub}_{X|{\rm Pa}(X)} \subset \mathcal{P}_{X|{\rm Pa}(X)}$, that is, to a strict subset of the usual possibilities.   We refer to such classical causal models as {\em parameter-restricted}  models for $G_i$, and denote them by \textsc{c}$_j G_i$, where $j$ is an index that ranges over the choice of parameter restriction.  More precisely, the $j$th parameter restriction is specified by specifying the set of allowed parameter values: $\mathcal{S}_j \equiv \{  \mathcal{P}^{\rm sub}_{X|{\rm Pa}(X)}\}_{X\in{\rm Nodes}(G_i)}$.

For a classical causal model based on a DAG $G_i$ and a parameter restriction described by $\mathcal{S}_j$, that is, for the model denoted by $\textsc{c}_jG_i$,
  the set $\mathcal{P}^{\textsc{c}_j G_i}$ of distributions on the set $O$ of observed variables in $G_i$ are those of the form
\beq
P_{O} = \sum_{{\rm Nodes}(G)/O} \left[ \prod_{X\in{\rm Nodes}(G)} P_{X|{\rm Pa}(X)} \right] 
\eeq
where $P_{X|{\rm Pa}(X)} \in \mathcal{P}^{\rm sub}_{X|{\rm Pa}(X)}$.   
In the case of the classical causal model for the DAG $G_i$  that is {\em parameter-unrestricted},  \textsc{c}$G_i$, the set of compatible distributions is denoted $ \mathcal{P}^{\textsc{c}G_i}$.

Clearly, the distributions that are compatible with a parameter-restricted classical causal model 
 \textsc{c}$_j G_i$, are included within those that are compatible with its parameter-unrestricted counterpart 
  $\textsc{c}G_i$,
$$\mathcal{P}^{\textsc{c}_j G_i} \subseteq \mathcal{P}^{\textsc{c}G_i}.$$
The interesting restrictions on the parameters, of course, are those for which we get a {\em strict} inclusion, $\mathcal{P}^{\textsc{c}_j G_i} \subset \mathcal{P}^{\textsc{c}G_i}$.


Let $\mathcal{P}^{\textsc{q}G}$ denote the set of distributions on the observed variables that are compatible with a parametrically quantum (hence parametrically radical) and structurally conservative model, assuming no restriction on the parameters.  

Since, in this article, we are only interested in classical causal models that can fit data consistent with operational quantum theory, we limit the scope of possible parameter restrictions to those that can achieve such a fit. More precisely, we consider only those parameter-restricted classical causal models, $\textsc{c}_j G_i$, for which there is sufficient parametric freedom that the compatible distributions subsume those predicted by operational quantum theory, i.e.,  those for which
 $\mathcal{P}^{\textsc{c}_j G_i} \supseteq \mathcal{P}^{\textsc{q}G}$.\footnote{In other words, we are not interested here in models $\textsc{c}_j G_i$ wherein $ \mathcal{P}^{\textsc{c}_j G_i} \subset \mathcal{P}^{\textsc{q}G}$  or wherein neither set subsumes the other because in both such cases one could rule out $\textsc{c}_j G_i$ relative to $\textsc{q}G$ in the standard way, by noting that it underfits the data relative to $\textsc{q}G$.}

  For these, it is useful to introduce a further distinction, based on whether or not the parametric freedom is such that the compatible distributions {\em go beyond} those of operational quantum theory: 
\begin{itemize}
\item[$\textsc{c}_j G_i$ is quantum-on-the-nose:] $ \mathcal{P}^{\textsc{c}_j G_i}= \mathcal{P}^{\textsc{q}G}$ 
\item[$\textsc{c}_j G_i$ is  quantum-extending:] $\mathcal{P}^{\textsc{c}_j G_i} \supset \mathcal{P}^{\textsc{q}G}$
\end{itemize}
In the quantum-on-the-nose case, the restriction on the set of parameters values is such that the set of compatible distributions on the observed variables for DAG $G_i$ coincides {\em precisely} with the set of distributions compatible with a quantum causal model for DAG $G$.
   In the quantum-extending case, the set of compatible distributions is a strict superset of the latter.

It is useful to make explicit the connection between the distinctions introduced here and the models discussed in the main text. For a Bell experiment, the DAG $G$ corresponding to the conservative causal structure 
 is that of Fig.~\ref{DAGs}(b) or (e), the common-cause structure, while the possibilities for DAGs $G_0, G_1,\dots$  corresponding to radical causal structures
  include the various different ways of allowing cause-effect relations between the labs (with the DAG of Fig.~\ref{DAGs}(c) being one way of doing so) and the various different ways of allowing for superdeterminism (with the DAG of Fig.~\ref{DAGs}(d) being one way of doing so).
 It follows that the model $\textsc{qCC}$ plays the role of $\textsc{q}G$, while $\textsc{cCE}_0$ and $\textsc{cSD}_0$ are examples of parameter-unrestricted $\textsc{c}G_i$ models.

With these distinctions in mind, we are now in a position to discuss the pertinence of our data-analysis technique to various existing attitudes towards causal accounts of Bell inequality violations.




\subsection{Reassessment of various loopholes in Bell experiments}

In discussions of Bell experiments, a ``loophole'' is generally taken to be a plausible reason for denying the validity of some assumption required for the derivation of the Bell inequalities (or at least the validity of this assumption for a particular experiment), such that one escapes the contradiction that exists between these assumptions and the observed violations of Bell inequalities. 


Consider a Bell experiment that violates a Bell inequality, but where the measurements on the wings are not done at space-like separation (such as the one described in this article). 
Such an experiment is said to suffer from a {\em locality loophole}, by which it is meant that one has a plausible reason for denying the validity of Bell's notion of local causality in such an experiment, namely, that positing a causal influence between the labs does not require those influences to be superluminal, 
so that such influences should be considered 
unobjectionable.\footnote{It is of course possible to object to allowing such a causal influence on the grounds that it is not mediated by any system that is described in the quantum formalism, but we shall not take this course here.}


For a Bell experiment that fails to achieve space-like separation, a model can {\em leverage} the locality loophole, that is, make use of a cause-effect relation between the labs, in order to reproduce the observations (in particular, the Bell inequality violations).
As a consequence, it is generally believed that 
 a good Bell experiment {\em must seal the locality loophole}, that is, it must enforce space-like separations between the measurements in the laboratories.

Recall that, as noted in the introduction, not underfitting the data is a low bar, and a better figure of merit in model selection is predictive power, which can be compromised not only by underfitting but by overfitting as well. 
Therefore, although leveraging the locality loophole allows one to ``reproduce the  observations'', i.e., to pass the low  bar of not underfitting the data, it is not clear a priori whether it allows one to pass the high bar of not {\em overfitting} the data.
Thus, even though positing a causal influence between the labs may be unobjectionable from the perspective of relativity theory, it can remain objectionable by virtue of having poor predictive power. 
 In this way, one can in principle provide evidence against the hypothesis of inter-lab causal influences using model selection techniques.

Note that the claim that it is critical to seal
 the locality loophole has not traditionally come with any caveats.  
For instance, no one has claimed that sealing this loophole is {\em only} important for causal models wherein the parameters are restricted but {\em not} important for causal models wherein the parameters are unrestricted.   
Consequently, we will take the claim that it is critical to seal the locality loophole to be overturned if it is overturned in the context of parameter-unrestricted causal models.
  Given that the parameter-unrestricted classical cause-effect model \textsc{cCE}$_0$ was found in our experiment to have less predictive power than the quantum common-cause model \textsc{qCC}, and given that all parameter-unrestricted models in the \textsc{cCE} class are likely to perform similarly to \textsc{cCE}$_0$ (as noted in our discussion section), our results show that one can indeed collect evidence {\em against the hypothesis of inter-lab causal influences}, without requiring the measurements to have been space-like separated.  





The {\em superdeterminism loophole} consists of the fact that it is possible to devise a superdeterministic model of a Bell experiment that reproduces the statistics predicted by quantum theory.  It is often claimed that this loophole cannot be tested experimentally.
As before, ``reproducing the statistics'' means merely not underfitting the data, while
%
 model selection techniques 
  hold models to a higher bar of predictive power, including {\em not overfitting} the data.    

Because the claim about the untestability of the superdeterminism loophole have not been predicated on restricting the parameter values in a superdeterminist model, we shall consider this claim to be overturned if it is overturned in the context of parameter-unrestricted models.  We have here shown that the parameter-unrestricted superdeterministic model \textsc{cSD}$_0$ has less predictive power than the quantum common-cause model \textsc{qCC}, and we have argued (in the Discussion section) that this is likely to be the case for all parameter-unrestricted superdeterministic models. 
Consequently, our results provide experimental evidence {\em against}
such superdeterministic models, contrary to the claim that the superdeterminist loophole cannot be tested.\footnote{See also Ref.~\cite{chaves2021causal} for a complementary perspective on how to modify a Bell experiment in order to be able to test the hypothesis of a  superdeterministic account of Bell inequality violations.}

\subsection{Models that are compatible with all and only the statistics predicted by quantum theory}


In a structurally radical model of the Bell experiment that is quantum-on-the-nose, the range of parameter values is so restricted that it is compatible with {\em all and only} the correlations achievable in the \textsc{qCC} model, and consequently satisfies the no-signalling condition for all parameter values in the restricted set.  In this case, one cannot hope to experimentally distinguish it from the \textsc{qCC} model.  
By choosing the parameter restriction carefully, one can in principle define a quantum-on-the-nose model for any structurally radical DAG that can reproduce Bell inequality violations, such as the DAGs associated to the \textsc{cCE}$_0$ and \textsc{cSD}$_0$ models.  

The standard view of Bohmian mechanics~\cite{sep-qm-bohm}
  is likely to be an example of such a model, where the DAG is one that allows for causal influences between the setting variable in one lab and the outcome variable in the other and so is in the \textsc{cCE} class.

We are not aware of any superdeterministic 
models that are quantum-on-the-nose, although in principle one could define models of this sort.  
Such a model would yield a novel type of superdeterminism 
 loophole insofar as it {\em would be} predicated on restrictions in the parameter values.  {\em This} loophole---unlike the one discussed above---{\em cannot} be closed by any experiment for which the correlations are consistent with operational quantum theory. 




\subsection{Models that are compatible with a superset of the statistics predicted by quantum theory}

Many proponents of structurally radical ways out of the Bell no-go result take their models to be {\em empirical competitors} to quantum theory, that is, they take them to be experimentally distinguishable from it in principle.  These structurally radical classical causal models are parameter-restricted but in a way that is quantum-extending.

We begin with those that are structurally radical by virtue of positing inter-lab causal influences.

We believe that Valentini's subquantum-nonequilibrium version of Bohmian mechanics~\cite{valentini1991signal1,valentini1991signal2,valentini2002signal,valentini2002signaldeterministic}, which can simulate quantum theory but also explicitly allows for violations of the no-signalling condition, is of this type. 

Because such models posit a more restricted scope of parameter values than we have assumed in the optimization that finds the best-fit parameter values, and because such restrictions could in principle {\em reduce} the degree of overfitting, 
the data analysis we have implemented here does not adjudicate between models of this sort and \textsc{qCC}.   Nonetheless, 
in order to be able to achieve such an adjudication using our technique, 
 it suffices 
to stipulate the range of parameter values that are allowed within a given model and to repeat our analysis while restricting the optimization  to this range.

Note also that although  one cannot conclude, based on our analysis, that \textsc{qCC} has more predictive power than a model of the sort just mentioned,
 it is still the case that our experiment provided an opportunity for finding that \textsc{qCC} has {\em less} predictive power than these alternatives.  That is, our experiment provided an opportunity for finding that nature exhibits deviations from operational quantum theory, for instance, by exhibiting the sorts of violations of the no-signalling condition that are predicted by Valentini's non-equilibrium version of Bohmian mechanics.  No evidence for such deviations was found however.  This is not unexpected insofar as such deviations are thought to only arise in exotic experimental scenarios, but a precision test in a non-exotic scenario nonetheless implies the {\em possibility} for finding such deviations. (See, e.g., the discussion of the distinction articulated in  Ref.~\cite{mazurek2021experimentally} between `terra-nova' and `precision' strategies for finding new physics.)  



Quantum-extending models that are superdeterministic  are also possible in principle.  (There has certainly been at least one proposal for a superdeterministic model that is empirically distinguishable from quantum theory, namely, that of Ref.~\cite{hossenfelder2020rethinking}, but it is unclear to us whether or not it subsumes all the statistics predicted by operational quantum theory.)
Our technique can in principle be used to adjudicate between such quantum-extending models and \textsc{qCC}.  It suffices to repeat our analysis with a fitting procedure that restricts the optimization to the appropriate range of parameter values.\footnote{Note that if a given proposed superdeterminist model {\em does not} subsume the statistics predicted by operational quantum theory, then it can also be disfavoured by our data analysis technique if our experimental data conforms with the quantum predictions. It is just that it will be disfavoured in the conventional way---on the grounds that it {\em underfits} the data.}

\subsection{Summary}

The conclusions we can draw  from our experimental results about the viability of various  causal models can be summarized as follows:
\begin{itemize}
\item[(i)] Our results provide evidence against the claim that it is possible to get around Bell's no-go result merely by endorsing 
one or another type of structural radicalism (together with parametric conservatism) 
without explicitly articulating any
 restriction on the scope of parameter values.
 Our results therefore provide a challenge to the position of those who suggest that Bell inequality violations imply superluminal influences or superdeterminism,
  but do not see fit to articulate any concrete model. 
\item[(ii)] Our results {\em do not} provide experimental evidence against parameter-restricted causal models that are quantum-on-the-nose (such as the standard version of Bohmian mechanics, or superdeterministic 
analogues thereof). 
\item[(iii)] Our results also do not provide experimental evidence against parameter-restricted causal models that are quantum-extending (such as Valentini's subquantum-nonequilibrium version of Bohmian mechanics~\cite{valentini1991signal1,valentini1991signal2,valentini2002signal,valentini2002signaldeterministic}, or superdeterministic 
analogues thereof), but our data analysis techniques {\em do} provide a means for experimentally assessing these against a quantum causal model if the parameter restriction that they endorse can be made precise.
\end{itemize}

\bibliography{Ref_2018}

\end{document}